\documentclass[aip,rsi,amsmath,amssymb,
 reprint,%
]{revtex4-1}

\usepackage{graphicx}
\usepackage{dcolumn}
\usepackage{bm}
\usepackage{amsmath}
\usepackage{amssymb}
\usepackage{color}
\renewcommand{\vec}[1]{\boldsymbol{#1} }

\begin{document}


\title{Radar for tracer particles}


\author{Felix Ott}%
\affiliation{ 
Experimentalphysik V, Universit\"at Bayreuth, 95440 Bayreuth, Germany}

\author{Stephan Herminghaus}
\affiliation{Max Planck Institute for Dynamics and Self-Organisation, 37077 G\"ottingen, Germany}

\author{Kai Huang}
 \email{kai.huang@uni-bayreuth.de}
\affiliation{ 
Experimentalphysik V, Universit\"at Bayreuth, 95440 Bayreuth, Germany}

\date{\today}

\begin{abstract}

We introduce a radar system capable of tracking a $5$\,mm spherical target continuously in three dimensions. The $10$\,GHz (X-band) radar system has a transmission power of $1$\,W and operates in the near field of the horn antennae. By comparing the phase shift of the electromagnetic wave traveling through the free space with an IQ-Mixer, we obtain the relative movement of the target with respect to the antennae. From the azimuth and inclination angles of the receiving antennae obtained in the calibration, we reconstruct the target trajectory in a three-dimensional Cartesian system. Finally, we test the tracking algorithm with target moving in circular as well as in pendulum motions, and discuss the capability of the radar system. 
\end{abstract}

							 



\maketitle

\section{\label{sec:intro}Introduction}

Since the pioneering work of Heinrich Hertz~\cite{Hertz1893} on the propagation of radio waves in free space, the development of radar (i.e., radio detecting and ranging) systems for remote sensing has benefited us in many different ways: From aircraft safety or traffic control, to weather prediction, as well as to space exploration~\cite{Skolnik2001}. In order to track an object that reflects electromagnetic waves, a radar system typically composes of electronic components that generate, transmit, receive, and process microwave signals~\cite{Skolnik2008}. Motivated by the widespread applications, a large variety of radar systems have been developed over the past century. Depending on the type of waveforms, it can be a pulse or continuous wave (CW) radar. Depending on the tracking algorithm, it can be a range or doppler radar. It may operate at a frequency as low as $\sim3$\,MHz for long range applications, or at a frequency as high as $\sim100$\,GHz for high spatial resolutions. The dimension of a radar system, which is largely determined by the size of the antenna array, ranges from a football-field sized long-range space-surveillance radar to a hand-hold one that measures the speed of an approaching baseball, or tracks the migration of insects~\cite{Skolnik2001,Oneal2004,Drake2012}. Based on the contrast in dielectric properties of water (rain) or ice particles (snow) to air, a meteorological radar monitors precipitation, cloud properties as well as the ocean topology at a global scale for weather forecasting~\cite{Atlas1990}. Considering the propagation of electromagnetic waves in soil, a ground penetration radar detects the variations of the dielectric constant underground, from which the compositions of the soil can be identified in three dimensions. This capability have contributed to geophysical investigations as well as to mining industries considerably~\cite{Skolnik2008}. In addition to object identification, a synthetic aperture radar provides high-resolution images of the target from a long distance~\cite{Sullivan2004}. To summarize, the development of radar technology in the past century provides us a high efficient and cost effective way of remote sensing and a substantial amount of knowledge on the relevant signal processing procedure. Taking these advantages, we here ask the following questions: How small can a target be tracked by a radar system? Can it be a millimeter sized tracer particle embedded in a granular medium?

The primary goal of introducing this particle tracking radar system is to understand the flow properties of granular materials (i.e., large agglomerations of macroscopic particles)~\cite{Jaeger1996}. Despite of their ubiquity in nature, industries and our daily lives, describing granular materials as a continuum, particularly in the vicinity of the transition from a solidlike to a liquidlike state, is still far from trivial~\cite{Duran2000}. With the help of computer simulations, we can predict to a certain extent the collective behaviors of granular materials, such as convection, agglomerations, silo blockage, mixing and segregations~\cite{pg13,Fortini2015,Huang2012}. At the particle level, a direct comparison of the predicted trajectories of computer simulations to experiments is always desirable to verify the model employed. However, the mobility of individual particles in a granular material, particularly in case of dense granular flow, is not easy to achieve, because the particles are optically opaque in most cases. 

Facing this challenge, a number of techniques have been employed to extract useful information at the particle level. They can be classified into the following three categories: (i) Techniques including magnetic resonance imaging (MRI)~\cite{Nakagawa1993,Ehrichs1995}, X-ray tomography~\cite{Neudecker2013,Athanassiadis2014}, laser sheet scanning~\cite{Toiya2007,Dijksman2012}, and confocal microscope~\cite{Weeks2000} provide full three-dimensional (3D) images of the sample with sufficiently high resolution to extract the positions of individual particles in the field of view. However, the limited frame rate due to the scanning time hinders the exploration of granular dynamics. Note that for the case of X-ray imaging, it is possible to bypass this obstacle by limiting the analysis to two-dimensional projections or using advanced synchrotron source~\cite{Royer2005,Maladen2009}. (ii) Spectroscopic approaches and autocorrelation functions in combination with videography can provide information on the averaged mobility of particles without identifying individual particles~\cite{Menon1997,Dixon2003,Fingerle2008,Huang2011}. (iii) Positron emission particle tracking (PEPT)~\cite{Parker1993,Parker1997} is capable of tracing a radioactive tracer particle by detecting the emitted gamma rays. More recently, the possibility of employing a radar system for particle tracking was also discussed~\cite{Hill2010}.   

Here, we introduce a continuous-wave (CW) radar system working at $10$\,GHz to track a small spherical particle continuously in 3D. It works in the near field of the antennae, and the target size is smaller than the wavelength of the electromagnetic wave $\lambda\approx3$\,cm being transmitted. In comparison to PEPT, the radar system is relatively compact, cost-effective and fits well into the normal laboratory conditions. Mover, it provides a continuous trajectory of the tracer particle in 3D with an high sampling rate, which is only limited by the analogue-digital converter.

\section{\label{sec:setup}Radar Setup}

\subsection{\label{sec:hw}Hardware}

\begin{figure}
\includegraphics[width = 0.45\textwidth]{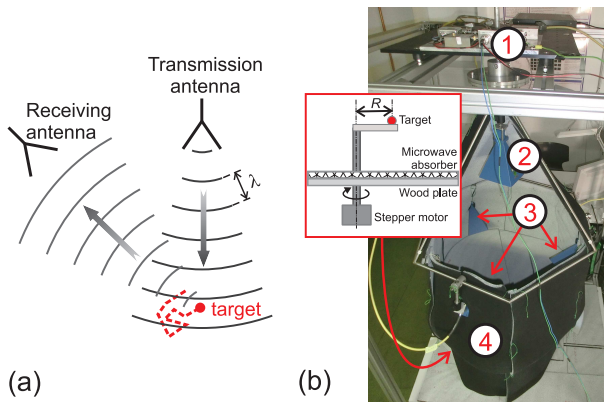} 
\caption{\label{fig:setup}(color online) (a) A sketch showing the tracking mechanism of the radar system. The wavelength of the emitted electromagnetic wave is $\lambda\approx3cm$. The target particle is a steel sphere with a radius of $5$\,mm. (b) A photograph of the experimental setup with the circuit board (1), transmission (2) and receiving (3) antennae, as well as the microwave absorbers (4) labeled. Inset of (b) shows a sketch of the target holder being connected to a stepper motor for system calibration.}
\end{figure}

As illustrated in Fig.\,\ref{fig:setup}\,(a), a $10$\,GHz electromagnetic wave is emitted by the transmission horn antenna (Dorado GH-90-20) into the free space, within which a spherical particle is moving in an unknown trajectory. The radar system obtains the target trajectory from the phase shift of received signal with respect to the emitted one. As the photograph of the setup (b) shows, all the four antennae are mounted in a hexagonal pyramid-shaped antenna holder with its base plane roughly parallel to the ground. The transmission antenna (label 2) mounted at the apex of the holder is pointing downwards toward the center of the hexagonal base. The three receiving antennae (label 3) are mounted symmetrically on the hexagonal base of the frame. For a better signal-to-noise ratio, all the antennae are adjusted to face the target region with the guide of a laser diode. The polarization of each antenna is also adjusted such that the output signal maximizes. In order to minimize the influence of the surrounding metallic objects, as well as to block the influence of other sources of electromagnetic radiation in the environment, microwave absorbers (Eccosorb AN-73, label 4) are used to cover the space where the target moves. The bottom of the space is closed by a horizontal wood plate covered with absorbers. The plate has a slit with adjustable position and width, through which a stepper-motor (Vexta PX243-02A) driven rotating arm is mount for the purpose of calibration [see the inset of (b)]. The rotation frequency, direction and the number of steps of the motor are controlled with a micro-controller (Arduino Uno). The circuit board of the system (label 1) emits, receives and compares electromagnetic waves. All circuit components are mounted on a thick metal plate, which serves as a heat sink. The whole system including the hexagonal pyramid-shaped antenna holder is mounted to an aluminum frame with shock absorbers.

\begin{figure}
\includegraphics[width = 0.45\textwidth]{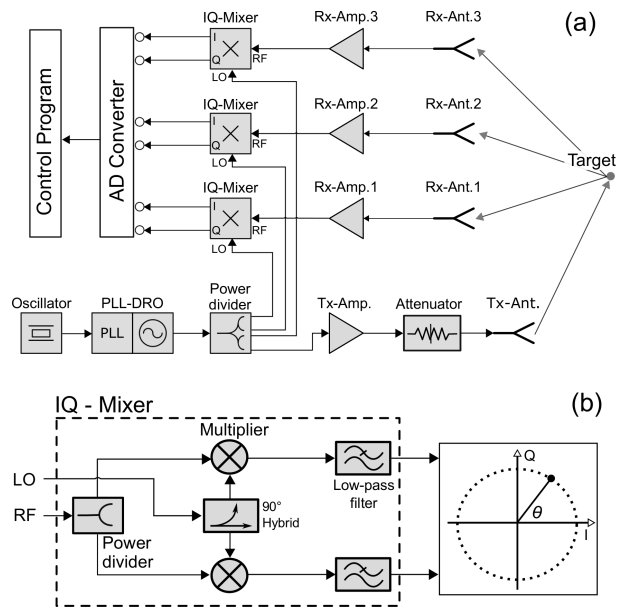} 
\caption{\label{fig:circ}(color online) Block diagram of the radar system (a) and that of an IQ-Mixer (left hand side of b). From the angle $\theta$ of the vector $I+Qi$ in a polar coordination system (right hand side of b), we obtain the relative phase shift of the electromagnetic wave traveling through the free space.}
\end{figure}

Figure~\ref{fig:circ}\,(a) shows the block diagram of the radar setup. The $10$\,GHz microwave signal with ultra-low phase noise is generated with PLL-DRO (phase-locked dielectric resonator oscillator, Miteq PLDRO-100-10000-15P). The reference frequency ($100$\,MHz) of the PLL-DRO is provided by a crystal oscillator (Miteq XTO-05-100-F-15P). The power of the generated signal is divided into four equal parts with a power divider (Miteq PD04-05001800). One of the divided signal is sent to the power amplifier of the transmission antenna (Tx-Amp., Miteq AMF-5B-09701020-33P-1) and each of the other three signals is sent to a IQ-Mixer (Miteq IRM0812LC2Q) as the local oscillator (LO) input. The power amplified signal, after proper damping with the attenuators (Aeroflex 18B-20 and 18B-10), is fed to the transmission antenna (Tx-Ant.). The horn transmission antenna shown in Fig.\,\ref{fig:setup}\,(b) has a pyramid shape with height $157$\,mm, length $140$\,mm, and width $104$\,mm. Linearly polarized electromagnetic waves are emitted off the transmission antenna to the target region which is roughly $1.2$\,m away. The scattered signal is detected by the three receiving antennae (Dorado GH-90-20), which are mounted roughly $0.7$\,m away from the target region. The signal obtained by each of the three receiving antennae, after power amplification (Rx-Amp., Miteq AFS4-09501050-09-S-4), is fed to the ratio frequency (RF) input of the corresponding IQ-Mixer. All circuit components are connected with each other with coaxial cables and a waveguide-to-coaxial adapter (Dorado WA90-S) is applied to each of the four antennae.

\subsection{\label{sec:algorithm}Tracking algorithm}

Figure~\ref{fig:circ}\,(b) illustrates the algorithm for an IQ-Mixer to detect the phase shift between the two input signals. Suppose the LO and RF input signals are $a\cos(2\pi f_{\rm 0}t)$ and $b\cos(2\pi ft + \theta)$, respectively. Here, $a$ and $b$ are the maximum amplitudes of the detected signals, $f_{\rm 0}$ and $f$ are the frequencies of the emitted and received electromagnetic waves. The RF signal is split into two equal parts. The quadrature hybrid component ($90^{\circ}$ Hybrid) splits the LO signal into two parts with one part being phase shifted by $90^{\circ}$ into $a\sin(2\pi f_{\rm 0}t)$. Subsequently, either part of the split signals is multiplied with one of the split RF signal, leading to

\begin{eqnarray}
V_{\rm I}=ab\cos(2\pi f_{\rm 0}t)\cos(2\pi ft + \theta), \nonumber \\ 
V_{\rm Q}=ab\sin(2\pi f_{\rm 0}t)\cos(2\pi ft + \theta). \label{eq:iq1}
\end{eqnarray}

\noindent As the high frequency components in Eqs.\ref{eq:iq1} are removed by the low pass filter, the output signals of the IQ-Mixer are

\begin{eqnarray}
I=\frac{ab}{2}\cos[2\pi(f_{\rm 0}-f)t - \theta], \nonumber \\
Q=\frac{ab}{2}\sin[2\pi(f_{\rm 0}-f)t - \theta]. \label{eq:iq2}
\end{eqnarray}

Because the target velocity is small in comparison to that of the electromagnetic wave, we extract the relative movement of the target from the phase shift of the vector $I+Qi$. The signals are fed to an analogue-digital (AD) converter (NI DAQPAD-6015) with a maximum sampling rate of $200$\,kHz and visualized in a complex plane with a Labview program. From the change of angle $\arctan(Q/I)$, we determine the movement of the target [see the illustration in Figure~\ref{fig:circ}(b)]~\cite{Sabah1998,Skolnik2008}. If the total path of the electromagnetic wave in free space varies with a distance of one wavelength, the vector $I+Qi$ rotates $2\pi$.

\subsection{\label{sec:radeq}Radar equation}

In this subsection, we discuss the sensitivity of the setup with the radar equation~\cite{Skolnik2001}

\begin{equation}
\label{eq:rad}
s=\left(\frac{P_{\rm Tx}G^2\lambda^2\sigma}{32\pi^2P_{\rm Rx}}\right)^{1/4}.
\end{equation}

\noindent It describes how the minimum working distance of the radar $s$ relies on the power of the transmitted electromagnetic wave $P_{\rm Tx}$, the gain of the antenna $G$, the radar cross section $\sigma=k\pi d^2/4$ with a factor $k$ and particle diameter $d$, and the minimum detectable power of the receiving antenna $P_{\rm Rx}$. Because the circumference of the sphere is comparable to the wavelength of the intercepted electromagnetic wave, the scattering is in the Mie region~\cite{Skolnik2001}. In this region, the factor $k$ fluctuates around $1$ with a maximum of $\approx 3$ at $\pi d/\lambda=1$. From Eq.\,\ref{eq:rad} and the radar cross section, we can estimate the minimum size of a particle being detectable by the system with

\begin{equation}
\label{eq:size}
d_{\rm c}=\frac{8s^2}{G\lambda}\sqrt{\frac{2\pi P_{\rm Rx}}{kP_{\rm Tx}}}.
\end{equation}

From the specifications of the microwave components, we have $G=20$, $P_{\rm Tx}=1$\,W (adjustable through varying the attenuator), and $P_{\rm Rx}\approx10^{-9}$\,W. If we choose $k=1$ and $s=2$\,m to be the total traveling distance of a electromagnetic wave\,\footnote{Note that the influence from the fluctuations of $k$ is relatively small because of the square root dependency.}, Eq.~\ref{eq:size} predicts a size of $\approx4.2$\,mm. Thus, a stainless steel sphere with $d=5$\,mm is chosen in the current investigation. A qualitative test with a particle of diameter $d=3$\,mm shows that amplitude $b$ of the received signal is too weak for an accurate determination of $\theta$.

\section{\label{sec:reconst}Coordinate transformation}

\begin{figure}
\includegraphics[width = 0.35\textwidth]{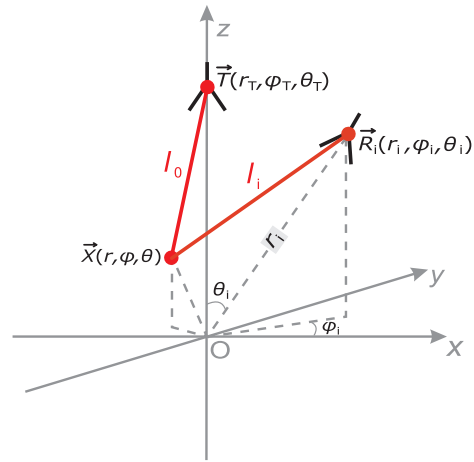} 
\caption{\label{fig:theo}(color online) The traveling path (red lines) of an electromagnetic wave in a Cartesian coordination system centered at $O$. $\vec{T}$, $\vec{R}$ and $\vec{X}$ are the positions of the transmission, receiving antennae as well as the target in the corresponding spherical coordination system.}
\end{figure}

The phase shift obtained from an IQ-Mixer due to target movement relies strongly on the exact location of the corresponding antenna. As shown in Fig.\,\ref{fig:theo}, the traveling distance of the electromagnetic wave in free space 

\begin{equation}
\label{eq:len0}
L_{\rm i}=l_{\rm 0}+l_{\rm i}
\end{equation}

\noindent is a sum of the distance from the target to the apex of the transmission antenna $l_{\rm 0}$ and that to the apex of the $i$th receiving antenna $l_{\rm i}$. As the phase shift has a limit of $2\pi$, what we obtain from the IQ-Mixer is only the modulo of $L_{\rm i}$ over $\lambda$. However, as the trajectory is continuous, we can obtain the relative movement of the target with respect to its initial position for a range greater than $\lambda$ by correcting the phase jump of $L_{\rm i} \pmod{\lambda}$. The goal is to obtain the position of the particle in a Cartesian coordinate centered at $O$. For the sake of simplify, we define the position of the transmission, the $i$th receiving antenna and the target to be $\vec T(r_{\rm T},\theta_{\rm T},\phi_{\rm T})$, $\vec R_{\rm i}(r_{\rm i},\theta_{\rm i},\phi_{\rm i})$, and $\vec X(r,\theta,\phi)$ in a spherical coordination system. Subsequently, we have

\begin{equation}
\label{eq:len}
l_{\rm 0}=r_{\rm T}\sqrt{1-2\cos{\alpha_{\rm T}}\frac{r}{r_{\rm T}}+{\left(\frac{r}{r_{\rm T}}\right)}^2}, 
\end{equation}
 
\noindent where $\alpha_{\rm T}$ is the angle between $\vec T$ and $\vec X$. If we choose $O$ to be the center of the target region, $r/{r_{\rm T}}$ is typically a small quantity. Thus, we can take the first order approximation and rewrite Eq.\,\ref{eq:len} as

\begin{equation}
\label{eq:l0}
l_{\rm 0}\approx r_{\rm T}-r\cos{\alpha_{\rm T}}. 
\end{equation}

\noindent This is equivalent to assume a plane wave propagation instead of a spherical one, because Eq.~\ref{eq:l0} represents the projected length of the vector $\vec T-\vec X$ on $\vec T$. As the main lobe of the emitted electromagnetic wave pattern has an angle of $\sim 15^{\circ}$, the phase error (the phase difference between a spherical wave and a plane wave) is relatively small and thus, the above assumption is appropriate. 

Similarly, we have for the receiving antenna

\begin{equation}
l_{\rm i}\approx r_{\rm i}-r\cos{\alpha_{\rm i}} \label{eq:li}
\end{equation}

\noindent with $\alpha_{\rm i}$ the angle between $\vec R_{\rm i}$ and $\vec X$. Because $\cos{\alpha_{\rm T}}=\vec T\cdot \vec X/(r_{\rm T}r)$ and $\cos{\alpha_{\rm i}}=\vec R_{\rm i}\cdot \vec X/(r_{\rm i}r)$, inserting Eqs.\,\ref{eq:l0} and \ref{eq:li} into Eq.\,\ref{eq:len0} yields

\begin{equation}
\label{eq:len1}
L_{\rm i}=r_{\rm T} + r_{\rm i} - \vec X(\vec e_{\rm T}+\vec e_{\rm i}),
\end{equation}

\noindent where $e_{\rm T}\equiv \vec T/r_{\rm T}$ and $e_{\rm i}\equiv \vec R_{\rm i}/r_{\rm i}$ are unit vectors pointing to the transmission as well as the $i$th receiving antenna. The above equation shows that the measured distance composes a constant $r_{\rm T} + r_{\rm i}$ that depends on the distances of the antennae to $O$ together with a small correction term (the center $O$ is chosen to be close to the target) that relies on the directions of the antennae. Thus, we need to determine the distance, azimuth as well as tilting angles of the antennae in order to reconstruct the trajectory of the target in a Cartesian system. Due to the modulo operation, one has to be aware that the system can only detect the relative movement of the target. Note that the directions here refers to the vectors $\vec e_{\rm T}$ and $\vec e_{\rm i}$, which are not exactly the same as the central axis of the antennae. The latter influences only the field of view of the radar system. 

\section{\label{sec:exp}Calibration}

In order to calibrate the system for the aforementioned parameters, we need to move the target in a pre-defined trajectory. As shown in the inset of Fig.\,\ref{fig:setup}(b), the target sphere is fixed on a rotating arm with a known distance to the rotating axis. If we define the center of the circular trajectory as $O$, and the direction of transmission antenna as the $z$ direction, we have $\vec e_{\rm T}=(0,0,1)$. Consequently, Eq.\,\ref{eq:len1} can be rewritten as

\begin{equation}
\label{eq:len2}
L_{\rm i}=r_{\rm T} + r_{\rm i} - \vec X\left(
\begin{array}{c}
\sin\theta_{\rm i}\cos\phi_{\rm i} \\
\sin\theta_{\rm i}\sin\phi_{\rm i} \\
1+\cos\theta_{\rm i} \\
\end{array} \right),
\end{equation}

\noindent where $\theta_{\rm i}$ and $\phi_{\rm i}$ are the tilting and azimuth angles of the $i$th antenna (see Fig.\,\ref{fig:theo}), respectively. Inserting the known trajectory of the target $\vec X=[R\cos(\omega t), R\sin(\omega t), 0]$ into the above equation, we have

\begin{equation}
\label{eq:len3}
\tilde{L_{\rm i}}=\tilde{r}_{\rm i} - R\sin\theta_{\rm i}\cos(\omega t+\phi_{\rm i}),
\end{equation}

\noindent where the angular velocity $\omega=10.312$\,s$^{-1}$ and the radius of rotation $R=6$\,cm. As described above, the IQ-Mixer only provides the relative distance $\tilde{L_{\rm i}}=L_{\rm i} - n\lambda$ instead of the absolute one, where $n=\lfloor L_{\rm i}/\lambda \rfloor$. Correspondingly, $\tilde{r}_{\rm i}=r_{\rm T} + r_{\rm i} - n\lambda$ is used in Eq.\,\ref{eq:len3}. It shows that the fluctuations of $\tilde{L_{\rm i}}$ are determined by the directions of the receiving antennae, or more specifically by $\theta_{\rm i}$ and $\phi_{\rm i}$, while the offset of the fluctuations is determined by the relative distances $\tilde{r}_{\rm i}$. Thus, we can obtain relevant parameters to reconstruct the target trajectory in the Cartesian coordinate through fitting the signal obtained from each channel with Eq.\,\ref{eq:len3}.

\begin{figure}
\includegraphics[width = 0.45\textwidth]{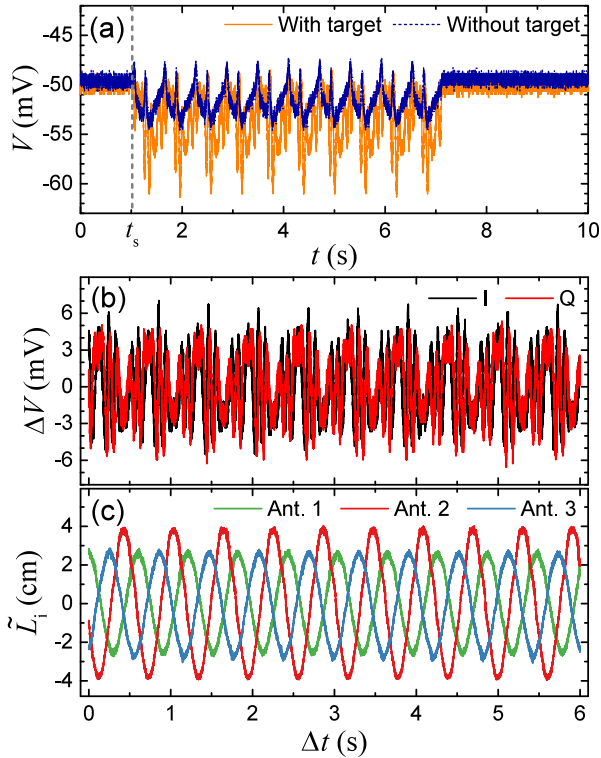} 
\caption{\label{fig:iqraw}(color online) Processing of the raw signals from the IQ-Mixers to obtain relative distances of the target with respect to the receiving antennae. The target circulates around the $z$ axis continuously for $10$ repetitions in the $x$-$y$ plane. (a) Sample voltage output (channel $2$, Q-signal) obtained with and without the target fixed to the rotating Styrofoam arm. The motor starts to rotate at time $t_{\rm s}$ with an angular velocity of $10.312$\,s$^{-1}$ and a radius of $6$\,cm. (b) Both I- and Q-signals of channel $2$ with background signals subtracted. (c) The relative distances with respect to the three receiving antennae obtained from the phase shift of each pair of I Q signals in a complex plane. $\Delta t$ corresponds to the relative time with respect to $t_{\rm s}$.}
\end{figure}

Figure~\ref{fig:iqraw} shows how to get $\tilde{L_{\rm i}}$ from the output signals of a IQ-Mixer. As shown in (a), the background signal obtained without the target fluctuates as the motor rotates, suggesting that the target holder is not completely transparent to the incoming electromagnetic waves. Note that even if the target holder is made of Styrofoam, a material with a dielectric constant very close to that of air, there still exist echo signals from the shaft and rotating arm of the target holder. As the whole process is repeated with the target, the signals obtained fluctuate with a larger amplitude as the motor rotates. As shown in (b), we obtain for each channel the I- and Q-signals induced by the target movement after subtract the background signal. Following the description in Sec.\,\ref{sec:algorithm}, we obtain the relative distance $\tilde{L}_{\rm i}=\lambda\cdot\arctan(Q/I)/(2\pi)$ from the angle of the vector $I+Qi$. After data smoothing with moving average, we see in (c) that $\tilde{L}_{\rm i}$ obtained from each channel fluctuates sinusoidally with an amplitude $\sim 3$\,cm. The smaller fluctuation amplitude with respect to $R$ is expected because the receiving antennae are facing the target with a tilting angle. A fit of the fluctuation of $\tilde{L}_{\rm i}$ with Eq.\,\ref{eq:len3} leads to the directions of each receiving antenna, which are shown in Table~\ref{tab:par}. For the sake of simplicity, we choose here the azimuth angle of the first receiving antenna as $0^{\circ}$.

\begin{table}
\caption{\label{tab:par}The parameters of the receiving antennae from the calibration.}
\begin{ruledtabular}
\begin{tabular}{lccr}
$i$ &$\theta_{\rm i}$&$\phi_{\rm i}$\\
\hline
1 & $25.0^{\circ}$ & $0^{\circ}$ \\
2 & $39.0^{\circ}$ & $252.6^{\circ}$ \\
3 & $26.2^{\circ}$ & $149.2^{\circ}$ \\
\end{tabular}
\end{ruledtabular}
\end{table}


In order to reconstruct the target trajectory from $\tilde{L}_{\rm i}$, we use 

\begin{equation}
\label{eq:reconst}
\left(
\begin{array}{c}
x \\
y \\
z
\end{array}
\right)=\frac{\vec{\tilde{r}}-\vec{\tilde{L}}}{\vec{T}},
\end{equation}

\noindent where $\vec{\tilde{r}}\equiv (\tilde{r}_{\rm 1}, \tilde{r}_{\rm 2}, \tilde{r}_{\rm 3})$, $\vec{\tilde{L}}\equiv (\tilde{L}_{\rm 1}, \tilde{L}_{\rm 2}, \tilde{L}_{\rm 3})$, and the transformation matrix is

\begin{equation}
\label{eq:trmat}
\vec{T} \equiv \left(
\begin{array}{ccc}
\sin{\theta_{\rm 1}}\cos{\phi_{\rm 1}} & \sin{\theta_{\rm 1}}\sin{\phi_{\rm 1}} & 1+\cos{\theta_{\rm 1}} \\
\sin{\theta_{\rm 2}}\cos{\phi_{\rm 2}} & \sin{\theta_{\rm 2}}\sin{\phi_{\rm 2}} & 1+\cos{\theta_{\rm 2}} \\
\sin{\theta_{\rm 3}}\cos{\phi_{\rm 3}} & \sin{\theta_{\rm 3}}\sin{\phi_{\rm 3}} & 1+\cos{\theta_{\rm 3}} 
\end{array}\right).
\end{equation}

According to Eq.\,\ref{eq:reconst}, $\vec{\tilde{r}}$, a constant vector that depends on the relative distances of the antennae with respect to $O$, contributes only an offset $\vec{\tilde{r}}/\vec{T}$ to the reconstructed trajectory. As the goal is to obtain the relative movement of the target, we set $\tilde{r}_{\rm i}=0$ so that the reconstructed trajectory always centered at $O$. Correspondingly, the offset of $\tilde{L}_{\rm i}$ is also removed before processing.

\section{\label{sec:circ}Target under circular motion }

\begin{figure}
\includegraphics[width = 0.45\textwidth]{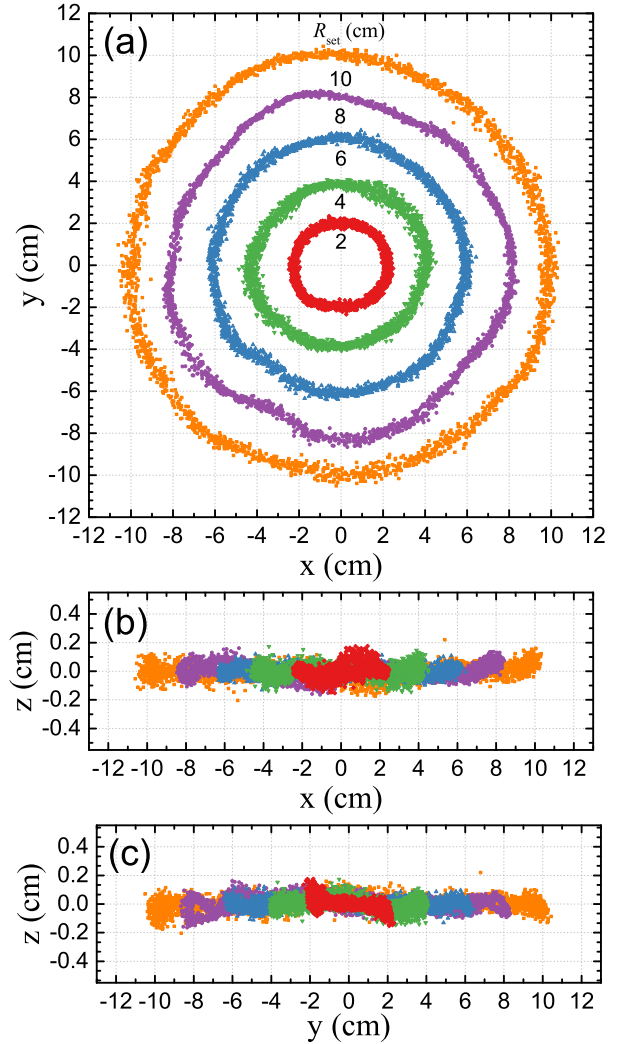} 
\caption{\label{fig:circle}(color online) Trajectories of the target projected to the $x$-$y$ (a), $x$-$z$ (b) and $y$-$z$ (c) planes of the Cartesian system. Five runs with different initially set radii of rotation $R_{\rm set}$ are performed. For each $R_{\rm set}$, the target circulates around the $z$ axis for $10$ repetitions.}
\end{figure}

Using the algorithm described above, we can reconstruct the trajectory of the target in a Cartesian system defined in the calibration process. More specifically, $O$ corresponds to the rotation center of the target, $z$ axis points to the apex of the transmission antenna, and $x$ axis is along the direction of first receiving antenna. As a first test, we trace the target rotating around the same axis, but with various radius $R_{\rm set}$ range from $2$\,cm to $10$\,cm. The upper limit of $R_{\rm set}$ is determined by the field of view of the radar system. The half-power ($-3$-dB) beam-width of the horn antennae used here is estimated to be $\sim 15$ degrees~\cite{carr2001}. For the transmission antenna mounted at a distance $r_{\rm T}\approx 1.2$\,m away from the target, we expect that the coverage area of main beam has a radius $\sim 16$\,cm. For the receiving antennae, which are $r_{\rm i}\approx 0.7$\,m away from the target region with a tilting angle $\theta_{\rm i}\approx 30^{\circ}$, we expect a coverage region with a radius $\sim 11$\,cm. 

Figure~\ref{fig:circle} shows the reconstructed trajectories of the target circulating around the $z$ axis for $10$ repetitions as in the calibration process. As demonstrated in (a), the radar system is capable of detecting the motion of the target and the trajectory transformation algorithm described above is appropriate. Quantitatively, the radius of rotation obtained from each reconstructed trajectory agrees with $R_{\rm set}$. In the radial direction, the data scattering has a standard deviation range from $0.13$\,cm to $0.29$\,cm, depending on $R_{\rm set}$. For the case of $R_{\rm set}=6$\,cm, the standard deviation $\sim0.18$\,cm is comparable to the error obtained by fitting $\tilde{L}_{\rm i}$ with a sinusoidal signal, which is $\sim 0.16$ for the antenna $1$ signal shown in Fig.\,\ref{fig:iqraw}(c). Possible sources of error are thermal noise~\cite{Skolnik2001}, mechanical noise of the stepper motor, as well as the mechanical stability of the antennae directions. For $R_{\rm set}=8$\,cm, the reconstructed trajectory distorts slightly from a circle, which presumably arises from the mechanical stability of the L-shaped target holder. Note that Styrofoam is a soft and foamy material susceptible to bending during rotation. The distortion is much less prominent for all other $R_{\rm set}$, suggesting that it should not arise from the coordinate transformation algorithm described above. As shown in (b) and (c), the data scattering in the $z$ direction has a range of $\sim 1$\,mm, which is much smaller than the size of the target $5$\,mm. Beside data scattering, there also exists a slight variation of $z$ with both $x$ and $y$ axis, or with the azimuth angle. This can also be attributed to mechanical stability of the target holder.

\begin{figure}
\includegraphics[width = 0.45\textwidth]{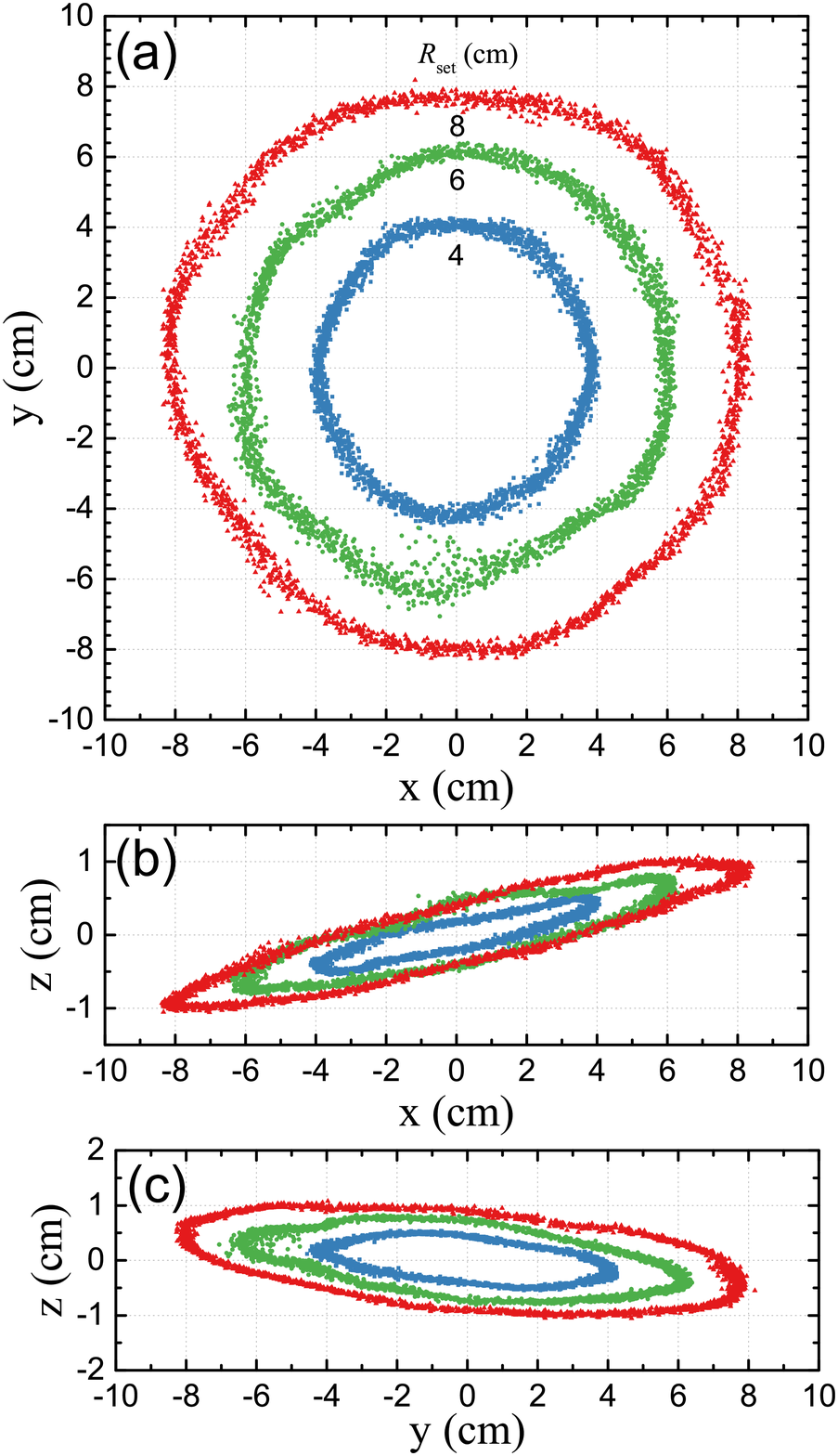} 
\caption{\label{fig:tilt}(color online) Reconstructed trajectories of the target circulating with $R_{\rm set}=4, 6$, and $8$\,cm, as the rotating axis is tilted by $\approx 7$ degrees with respect to the $z$ axis. Other configurations are the same as in Fig.\,\ref{fig:circle}.}
\end{figure}

Stepping further, we tilt the rotating axis of the motor slightly $\sim 8^{\circ}$ and repeat the above analysis with $R_{\rm set}=4$, $6$ and $8$\,cm. Note that the coordination system is fixed during the calibration process, the reconstructed trajectories should also tilt correspondingly. Figure~\ref{fig:tilt} shows the reconstructed trajectories of the target. As shown in (b) and (c), the tilting of the trajectories for all $R_{\rm set}$ is clearly distinguishable. More specifically, the projection of the new rotating axis on the $x$-$y$ plane points to the $2$nd quadrant. The data scattering is on the same order as in Fig.\,\ref{fig:circle}. Because of the small titling angle, the ellipticity of the trajectories projected to the $x-y$ plane (a) is less obvious in comparison to (b) and (c). The tilting angle $7.1\pm0.6^{\circ}$ is measured via transforming the reconstructed trajectories into the corresponding spherical coordinate and averaging over the elevation angles of all data points. Subsequently, an average over the distances of all data points of each trajectory to $O$ gives rise to the measured radius of the circular motion.

\begin{figure}
\includegraphics[width = 0.45\textwidth]{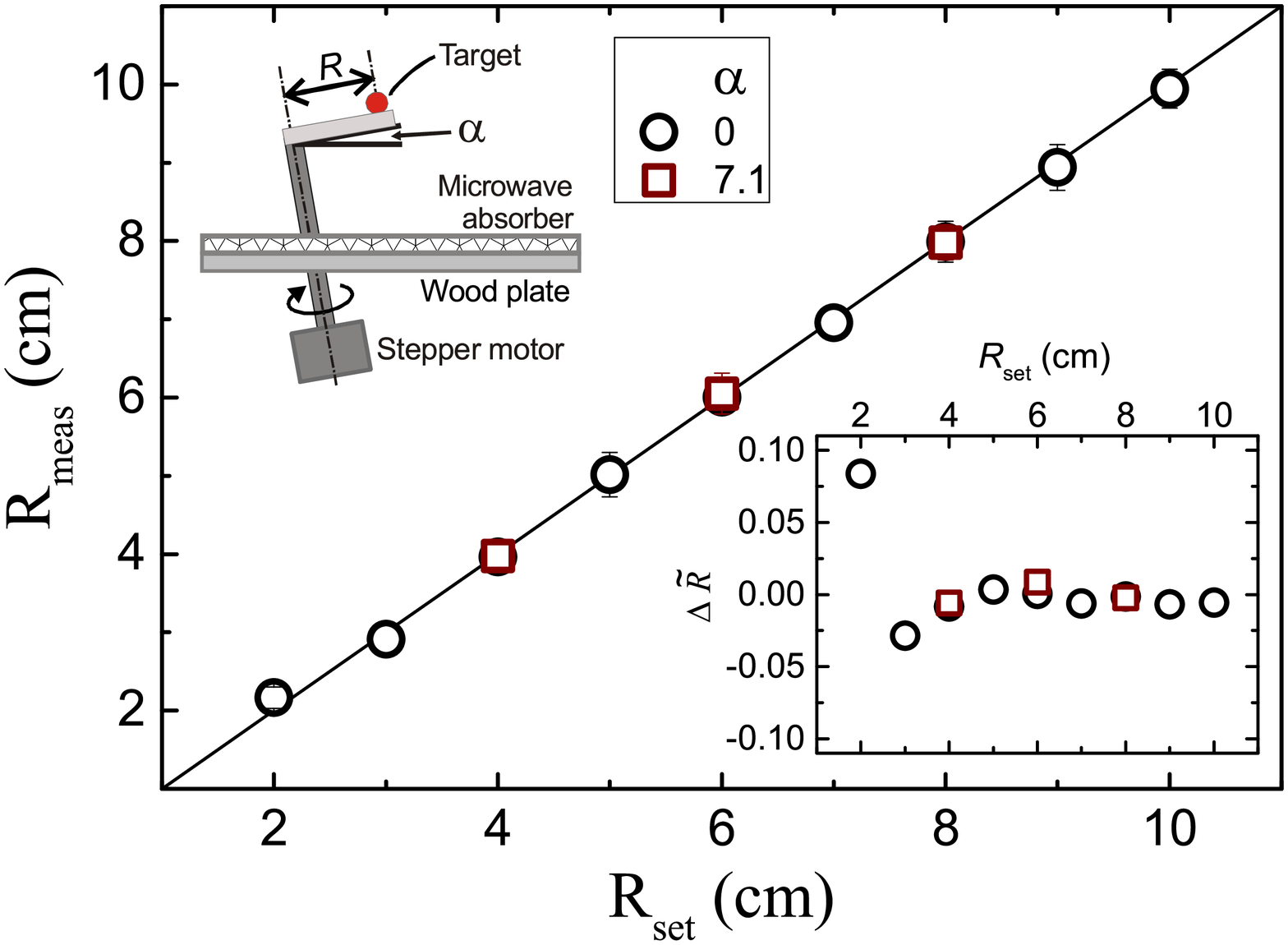} 
\caption{\label{fig:rcmp}(color online) A comparison of the radius of the target trajectory obtained from the radar setup $R_{\rm meau}$ and the one initially set $R_{\rm set}$. Inset shows the relative difference $\Delta\tilde{R}=(R_{\rm meau}-R_{\rm set})/R_{\rm set}$ as a function of $R_{\rm set}$.}
\end{figure}

Finally, we compare the radius of the circulating target obtained from each reconstructed trajectory with the one initially set while fixing the target to its holder. As shown in Fig.\,\ref{fig:rcmp}, the agreement is reasonably well for both non-tilted and tilted cases. Such an agreement indicates that, following the calibration and signal process protocol described above, the radar system is capable of tracking a spherical target of $5$\,mm in diameter with a reasonably good accuracy. The uncertainty of the measured radius is for most cases within the size of the symbol, as the data scattering is relatively small. In the inset of Fig.\,\ref{fig:rcmp}, we show the relative difference of the measured radius with respect to $R_{\rm set}$. It shows that, except for the smallest $R_{\rm set}$, the relative error is within $5\%$. For $R_{\rm set}=2$\,cm and $\alpha=0$, the measured radius is $\sim 13\%$ larger than $R_{\rm set}$. This difference can be attributed to the uncertainty of $R_{\rm set}$: As there is a fixed uncertainty for fixing the target on its holder, the smaller the  $R_{\rm set}$, the larger the relative error. 

\section{\label{sec:swing}Target under swinging motion}

\begin{figure}
\includegraphics[width = 0.45\textwidth]{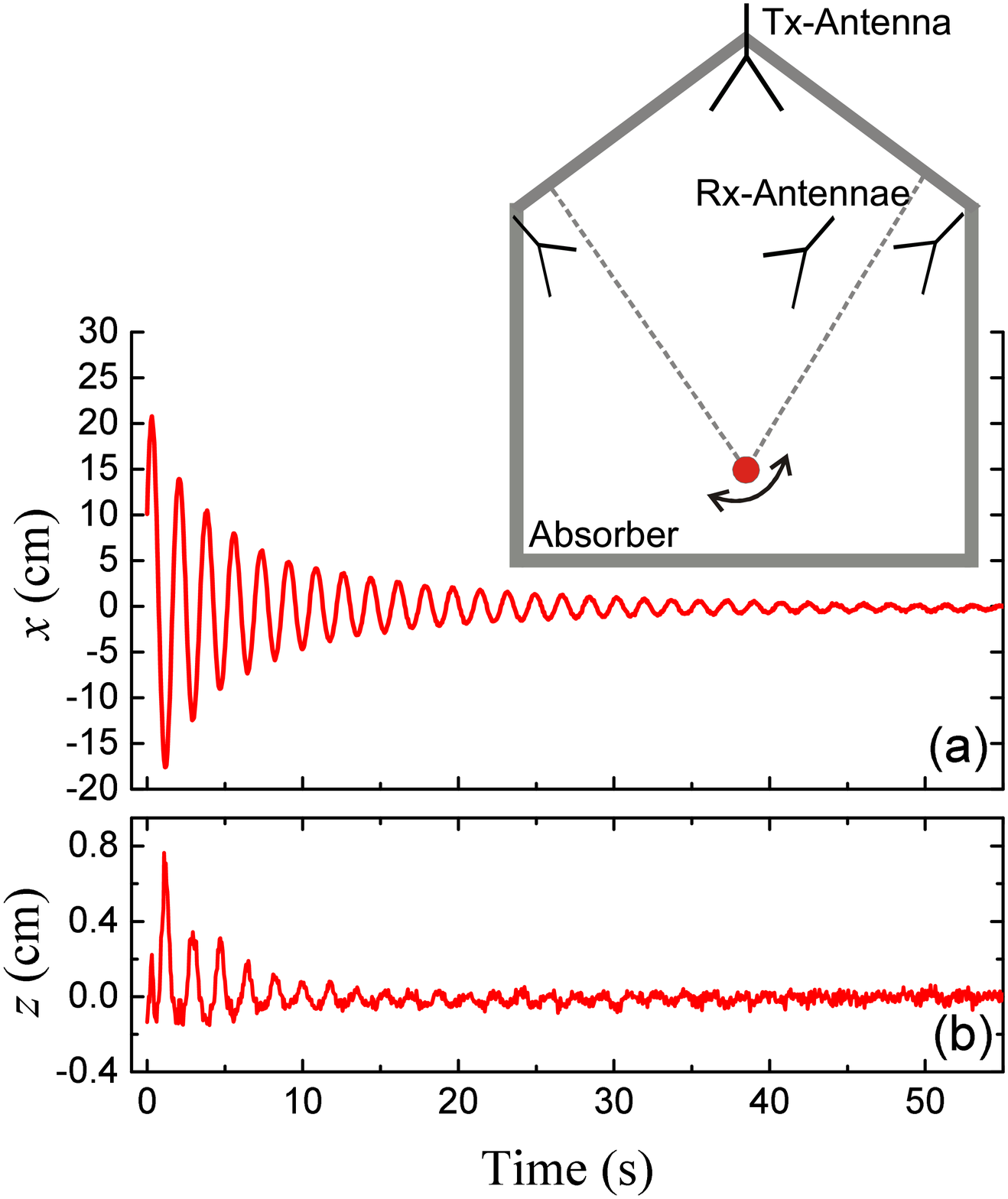} 
\caption{\label{fig:swing}(color online) The reconstructed trajectory of the target undergoing a pendulum motion, as sketched in the inset of (a). (a) and (b) correspond to the oscillations of the trajectory in the $x$ and $z$ axis, respectively. Here the $x$ axis represents the oscillation direction of the trajectory projected to the $x-y$ plane.}
\end{figure}

Beside circular motion, we also suspend the target with two thin rigid strings to create a pendulum [see the inset of Fig.\,\ref{fig:swing}\,(a)]. From the reconstructed trajectory of the swing motion, we obtain the oscillation period of the pendulum and compares it with the predicted value $1.78\pm0.01$\,s, which is estimated from the measured length of the pendulum $L\approx 78$\,cm. Here, we use two strings instead of one for the purpose of confining the pendulum motion into a vertical plane. The target is suspended to a fixed frame on which the antennae are mounted; that is, a fixed coordination system. The advantage of this configuration in comparison to the motor driven circular motion is that the unnecessary echoes of the target holder and the influence of the rotating axis are avoided.

Figure~\ref{fig:swing} shows the reconstructed trajectory of the target as a damped oscillator. For a better characterization of the oscillation amplitude, the coordinate presented here, in comparison to the one defined during the calibration process, is rotated along the $z$ axis such that the oscillation is along the $x$ axis. As shown in (a), the initial amplitude of oscillation is $\sim 20$\,cm, suggesting that even if the target moves away from the beam-width of the receiving antennae (i.e., the expected field of view), it is still possible for the receiving antennae to sense its movement. Note that the radius of the field of view is estimated with $r_{\rm i}\tan{\psi}/\cos{\theta_{\rm i}}\approx 11$\,cm (see discussions at the beginning of Sec.\,\ref{sec:circ}), where the typical values of the distance to the receiving antenna $r_{\rm i}=0.7$\,m, half beam width $\psi\approx 8^{\circ}$ and the tilting angle of the receiving antennae $\theta_{\rm i}\approx 30^{\circ}$ are used. From the reconstructed trajectory, we obtain an averaged oscillation period $1.76\pm0.07$\,s, which agrees with the predicted value within the error. The corresponding oscillation in the $z$ direction is shown in (b). In principle, either positive or negative peaks in the $x$ direction leads to a positive peak in $z$ direction, provided that the symmetric axis (i.e., the direction of gravity) is along the $z$ direction. However, a comparison of the fluctuations between (a) and (b) reveals that the peaks corresponding to the positive value of $x$ are suppressed. See, for example, the first peak in the $z$ direction is lower than the second one. It suggests that the $z$ axis is tilted with respect to the direction of gravity.

\section{\label{sec:err}Discussions on the capability of the radar system}

In the above analysis, we have demonstrated that the downsized radar system working in the near field of the antennae can track a spherical particle with a diameter smaller than the wavelength. The $10$\,GHz radar system operates at a transmission power of $1$\,W and detects echo signals with a minimum power of $10^{-9}$\,W. Note that, for a typical radar system, the detectable echo signal power is $10^{-13}$\,W, which is orders of magnitude smaller than the current system used~\cite{Skolnik2008}. Thus, based on Eq.\,\ref{eq:size}, it is possible to track even smaller particles  with a more sensitive antenna. However, we need to notice that an increase of the antenna gain (i.e., more sensitive antenna) may lead to a dramatic increase of the antenna dimensions. Due to the scaling with $s^{4/3}$, an efficient way of tracking a smaller particle is to put it closer to the antennae, although the field of view is sacrificed. Another convenient approach is to decrease $\lambda$ to millimeter range (i.e., to use higher frequency), which results in a higher cost of the system. In addition, one has to consider the attenuation of the atmosphere if millimeter waves are implemented~\cite{Skolnik2008}. 

Concerning the sampling rate, the radar system introduced here has an analogue signal output from the IQ-Mixer. In principle, it provides a continuous trajectory of the target in three dimensions. In reality, the sampling rate $200$\,kHz is limited by the AD converter, which is introduced for the sake of digital signal processing. Thus, the maximum speed of the tracer particle can be much higher in comparison to a PEPT system~\cite{Parker1993, Parker1997}, which is $\approx2$\,ms$^{-1}$. As the reconstruction algorithm does not require much computing power, it is also possible to track the target in real time as most of the radar systems do. As the field of view is only limited by the power of the echo signal, it is possible to tune the antenna direction based on the instantaneously detected target locations in order to expand the field of view. Nevertheless, this approach requires the change of antenna directions to be be monitored accurately. 

\section{Conclusions}

To summarize, we demonstrate the possibility of tracking a spherical particles with a diameter of $5$\,mm in real time and in three dimensions using a radar system operating at a frequency of $10$\,GHz. The size limit of the target can be understood by the radar function, which suggests the possibility of tracing even smaller particles by modifying the emitted power of the transmission antenna (currently $1$\,W), the working distance as well as the frequency of the electromagnetic wave. The system can be calibrated with the target moving in a well defined trajectory. Based on the directions of the antennae obtained from the calibration process, the trajectory of the target in a Cartesian system can be reconstructed. Test runs with the target moving in circular trajectories as well as in a pendulum motion have been performed. The reconstructed trajectories of the target from the radar system agree with the known parameters of the trajectories, demonstrating the capability of the particle tracking radar. The field of view of the radar system, which relies on the beam width and the distance of the antenna to the target, covers at least a region of 12x12x12cm$^3$. 

In the future, it is essential to characterize further possible sources of noise in order to enhance the signal-to-noise ratio for more accurate positioning. As the Styrofoam target holder with a low dielectric constant used in the first test doesn't hinder the detection of the target, it is possible to track the target even if it is embedded in a granular material. More investigations for embedded situations are necessary before applying the radar system to granular flow problems. Last but not least, the possibility of tracking multiple targets, as in an automatic detection and tracking (ADT) radar system~\cite{Skolnik2008}, should also be addressed.

\begin{acknowledgments}

The authors acknowledge the technical support from Udo Kraft, Klaus Oetter and Michael Rozmann. This work is partly supported by the German Research Foundation through Grant No.~HU1939/4-1. 



\end{acknowledgments}


\nocite{*}
%

\end{document}